\documentclass[letterpaper,11pt]{article}   
\usepackage{LeGquant-ph,amssymb, amsmath, graphicx, legall-macro,mathrsfs}
\begin{document}   

\vspace{14mm}
\begin{center} 
{\LARGE \bf Quantum Weakly Nondeterministic }\vspace{2mm}

{\LARGE \bf Communication Complexity}\vspace{8mm}  

{\large Fran{\c c}ois Le Gall}\vspace{6mm}

{\it Department of Computer Science, The University of Tokyo}\\
{\it 7-3-1 Hongo, Bunkyo-ku, Tokyo 113-0033, Japan} \vspace{1mm}

and\vspace{1mm}   

{\it ERATO-SORST Quantum Computation and Information Project, JST}\\
{\it Hongo White Building, 5-28-3 Hongo, Bunkyo-ku, Tokyo 113-0033, Japan}\vspace{3mm}

email: legall@qci.jst.jp\vspace{8mm}

\setlength{\baselineskip}{9.5pt}
      \begin{quotation}
\noindent{\bf Abstract.}\hbox to 0.5\parindent{}
In this paper, we study a weak version of quantum nondeterministic communication complexity, corresponding to the
most natural generalization of classical nondeterminism, in which
a classical proof has to be checked with probability one by a quantum protocol.
Another stronger definition of quantum nondeterminism has already been extensively studied,
corresponding to the view of quantum nondeterminism as unbounded-error one-sided quantum computation, 
but, although being mathematically convenient, this definition fundamentally lacks the original view of nondeterministic processes 
as proof-checking processes. 
In this paper, we prove that, in the framework of communication complexity, 
even the weak version of quantum nondeterminism is strictly stronger than classical nondeterminism.
More precisely, we show the first separation, for a total function,
of quantum weakly nondeterministic and classical nondeterministic communication complexity.
This separation is quadratic and shows than classical proofs can be checked more efficiently by
quantum protocols than by classical ones, in the framework of communication complexity.
\end{quotation}
\end{center} 

\section{Introduction}   
\subsection{Quantum nondeterminism}
Classical nondeterminism, although being an unrealistic model
of computation, is a fundamental concept
in computational complexity with practical applications, 
as shown, for example, by the importance of the theory of $NP$-completeness.
There are two different views of classical nondeterminism.
A nondeterministic process computing a Boolean function $f(x)$
can be seen as a deterministic process $B$ receiving, besides the input $x$, a guess, or proof, $z$ and
satisfying  
the following conditions:
If $f(x)=1$ there should exist a proof $z$ such that $B(x,z)=1$; if $f(x)=0$ then
$B(x,z)=0$ for all proofs $z$.
Another view of nondeterminism is to consider $B$ receiving no proof, but being probabilistic.
Then $B$ should output $1$ with positive probability if and only if $f(x)=1$.
It is easy to see that the two models are perfectly equivalent in the classical setting. 

These two views of nondeterminism have been extended to obtain two alternative definitions of quantum 
nondeterminism.
The first one, that we call in this paper quantum strong nondeterminism, 
is the quantum version of the probabilistic view of nondeterminism:
the quantum process $B$ should output $1$ with positive probability if and only if $f(x)=1$.
The second one, that we call quantum weak nondeterminism, is the extension of the first view of nondeterminism:
If $f(x)=1$ there should exist a classical proof $z$ such that $B(x,z)=1$ with probability 1; 
if $f(x)=0$ then $B(x,z)=0$ with probability $1$ for all classical proofs $z$. In this case,
$B$ is thus an exact quantum checking procedure.
The point is that, contrary to the classical case,
in the quantum setting these two definitions do not seem equivalent and, in the query complexity framework,
strong nondeterminism has be shown to be indeed stronger than weak 
nondeterminism: de Wolf \cite{deWolfSIAMJC03} has provided a total function for which the strongly quantum nondeterministic query complexity is $O(1)$, 
while its quantum weakly nondeterministic query complexity is $\Omega(\sqrt{n})$, where $n$ is the input length.

The main advantages of the strong version of quantum nondeterminism is that the definition
is mathematically very convenient and that it leads to many interesting results.
For quantum Turing machines, this gives a complexity class known as
quantum-$NP$, which has been shown to be equal to the classical complexity class $co-C_{=}P$ \cite{Yamakami+99}. For communication protocols,
de Wolf \cite{deWolfSIAMJC03} has presented an algebraic characterization of quantum
strongly nondeterministic communication complexity. Moreover, unbounded ($O(1)$ vs.~$\Omega(\log n))$)
and exponential ($O(\log n)$ vs.~$\Omega(n)$) gaps are known between
quantum strongly nondeterministic and classical nondeterministic
communication complexity of some total functions. 
The latter results show the
power of quantum strong nondeterminism but, in our opinion, this concept
is in a way too powerful to be directly compared with classical nondeterminism.
Above all, it lacks the view of nondeterminism as a proof that
can be efficiently checked, a view that has been fundamental in complexity theory, for example leading
to concepts such as probabilistically checkable proofs (PCP). We refer to \cite{deWolfSIAMJC03} for another discussion about
these two definitions and a third natural definition where the proof is allowed to be a quantum state, that we will not consider
in this paper. 
We only mention that, 
although quantum proofs can be extremely useful in some cases 
(see in particular the works \cite{AaronsonCCC03,Raz+CCC04} studying the power of
quantum proofs in certificate complexity and communication complexity, but in the setting where proofs have to be checked only with high probability), 
as far as quantum weakly nondeterminism is concerned, the proof has to be 
checked without error and, in this case, the advantage of quantum proofs over classical proofs is not obvious at all.

\subsection{Our contributions}
In this paper, we focus on quantum weak nondeterminism and particularly
quantum weakly nondeterministic communication complexity, which has, to our knowledge, never been studied before this work. 
We show 
a quadratic gap between classical nondeterministic and quantum weakly nondeterministic 
communication complexity for a total function. 
We believe that this separation of classical nondeterministic communication complexity
and the weakest model of quantum nondeterministic communication complexity, although being
only quadratic, is another indication of the power of quantum computation. Indeed, the proof being classical, such a separation reveals that, 
if quantum exact checking procedures are allowed, the process of guessing proofs is more powerful than with classical deterministic checking
procedures.

Many separations of quantum and classical communication complexity 
are known in the usual two-players model \cite{Aaronson+05,Bar-Yossef+STOC04,Buhrman+STOC98,Hoyer+STACS02,KlauckSTOC00,RazSTOC99,deWolfSIAMJC03}.
In particular, an exponential separation of quantum exact communication complexity and classical nondeterministic 
communication complexity has been shown for a partial function (i.~e.~a function where the inputs satisfies a promise) by 
Buhrman, Cleve and Wigderson \cite{Buhrman+STOC98}. 
But, except de Wolf's result \cite{deWolfSIAMJC03}, no gap larger than quadratic between classical
and quantum complexity, for any mode of computation, is known for total functions. 
Moreover, before the present work, the polynomial separations for total functions
already found \cite{Aaronson+05,Hoyer+STACS02,Buhrman+STOC98,KlauckSTOC00} were based on database search-like problems, 
that are trivial if classical nondeterminism is allowed, and thus cannot
be used to show a gap between quantum weak nondeterminism and classical nondeterminism.
The total function we consider in order to show the separation is new, based of the concept of Hadamard codes, and is inspired by
a function considered by Buhrman, Fortnow, Newman and R\"ohrig \cite{Buhrman+SODA03}
in the slightly different framework of query complexity and property testing. 

We present an efficient quantum weakly nondeterministic
protocol computing our function, that generalizes the protocol in \cite{Buhrman+SODA03},
based on the local testability property of Hadamard codes and the fact that,
with the promise that a string is in the Hadamard code, the string can be decoded efficiently
using Bernstein-Vazirani algorithm \cite{Bernstein+SIAMJC97}. 
The main contribution of our work is the proof of a classical lower bound on the number of bits of communication necessary for a classical
nondeterministic protocol, obtained by showing an upper bound on the number of inputs for which each message can be used, which
is basically a problem of extremal combinatorics.
Proving this upper bound is indeed the hard part of the proof. 
This gives a separation $O(\log n)$ vs.~$\Omega(\log^2 n)$, where $n$ is the input length,
of respectively quantum weakly nondeterministic and classical nondeterministic communication 
complexity, for our total function.

The paper is structured as follows. 
We present definitions 
in Section \ref{section:notations}.
We then show the quantum upper bound in Section \ref{section:upperbound} and the classical lower bound in 
Section \ref{section:lowerbound}. Finally, in Section \ref{section:open}, we discuss open problems. 

\section{Notations and Definitions}\label{section:notations}
\subsection{Notations}
In this paper, we will mainly work in vector spaces of the form $\{0,1\}^n$ with the usual addition between
vectors $\mathbf{x}=(x_0,\ldots,x_{n-1})$ and $\mathbf{y}=(y_0,\ldots,y_{n-1})$ defined as
$\mathbf{x}\mathbf\oplus\mathbf{y}=(x_0\oplus y_0,\ldots,x_{n-1}\oplus y_{n-1})$,
where $x_i\oplus y_i$ denotes the parity of $x_i$ and $y_i$, and the inner product defined as 
$\mathbf{x}\cdot \mathbf{y}=\bigoplus_{i=0}^{n-1} x_iy_i$.
We will in several occasions consider integers in $\Set{0}{2^n-1}$ as vectors of $\{0,1\}^n$ through their binary encoding.

We define the function $\delta$ over $\Integers\times\Integers$ as follows.
$$\delta(a,b)=\left\{\begin{array}{ll}
0&\textrm{ if } a=b\\
1& \textrm{ if } a\neq b
\end{array}
\right.,
\textrm{    for any integers } a \textrm{ and } b.$$
For $k\ge 1$, we denote by $S_k$ the set $\Set{1}{2^k-1}\backslash\{2^j\:\vert\:0\le j\le k-1\}$, i.~e.~the set of integers in 
$\Set{1}{2^k-1}$ that are not a power of $2$. Finally, for any $i\in\Set{1}{2^k-1}$, we denote by $[i]$ the larger
power of $2$ smaller or equal to $i$. In other words, $[i]=2^{\floor{\log_2 i}}$.

We now recall the definition of Hadamard codes.
\begin{definition}
For any integer $k\ge 1$, the Hadamard code of length $2^k$, denoted $\mathscr{H}_k$, is the set
$$\big
\{h(\mathbf{w})\:\vert\:\mathbf{w}\in\{0,1\}^k\big\},
$$
where $h(\mathbf{w})$ is the binary vector of length $2^k$ with $i$-th coordinate $\mathbf{w}\cdot \mathbf{i}$ $($for $\:0\le i\le 2^k-1)$.
\end{definition}
Notice that $\mathscr{H}_k$ is a linear code containing $2^{k}$ codewords of length $2^k$.

\subsection{Nondeterministic communication complexity}\label{section:comcom}
\subsubsection{Classical nondeterministic protocols}
We first recall the definition of classical nondeterministic communication complexity.
We refer to the textbook by Hromkovi{\v c} \cite{Hromkovic97}
for further details.
Given a set of pairs of strings $X\times Y$, where $X\subseteq\{0,1\}^\ast$ and
$Y\subseteq\{0,1\}^\ast$, and a function $f:X\times Y \to \{0,1\}$,
the communication problem associated to $f$ is the following: 
Alice has an input $x\in X$, Bob an input $y\in Y$ and their goal is to compute
the value $f(x,y)$. We suppose that Alice and Bob have unlimited computation power.
Moreover, a proof is given to the protocol: Alice and Bob each receive a string which is private, i.~e.~each player cannot see
the other's part of the proof.
We say that a protocol $P$ is a nondeterministic protocol for $f$ if, for each $(x,y)\in X\times Y$, the following holds:
\begin{enumerate}
\item[(i)]
if $f(x,y)=1$ then there is a proof such that the protocol outputs $1$, 
\item[(ii)]
if $f(x,y)=0$ then, for all proofs, the protocol outputs $0$.
\end{enumerate} 
The communication complexity of a nondeterministic protocol $P$ that computes correctly $f$, 
denoted $N(P,f)$, is the maximum, over all
the inputs $(x,y)$ and the proofs, of the number of bits exchanged between Alice and Bob on this input.
The nondeterministic communication complexity of the function $f$, denoted $N(f)$, is the minimum,
over all the nondeterministic protocols $P$ that compute $f$, of $N(P,f)$.

We now recall the notions of rectangle, covering and their relation with classical nondeterministic complexity.
A rectangle of $X\times Y$ is a subset $R\subseteq X\times Y$ such that $R$ can be written as $A\times B$
for some $A\subseteq X$ and $B\subseteq Y$. The rectangle $R$ is said to be 
1-monochromatic for $f$ if, for all $(x,y)\in R$, $f(x,y)=1$. 
A 1-covering of size $t$ for $f$ is a set of $t$ 
rectangles $R_1,\cdots,R_t$ of $X\times Y$ that are 1-monochromatic for $f$ and
such that $R_1\cup\cdots\cup R_t=\{(x,y)\in X\times Y\:\vert\: f(x,y)=1\}$.
Let $C^1(f)$ be the minimum, over all the 1-covering of $f$, of the size of the covering. 
Then the following fact holds (we refer to \cite{Hromkovic97} for the proof).
\begin{fact}\label{fact1}
$N(f)=\ceil{\log_2{C^1(f)}}$.
\end{fact} 

\subsubsection{Quantum weakly nondeterministic protocols}
Let us now consider quantum communication complexity.
We refer to Nielsen and Chuang \cite{Nielsen+00} for details about quantum computation
and to \cite{Buhrman00,Klauck00,deWolfTCS02} for good surveys of quantum communication complexity.

We define quantum weakly nondeterministic protocols as in the classical case, the only modification being that the messages are now allowed to be
quantum: Alice and Bob receive inputs $x$, $y$ and two classical strings corresponding to a classical proof,
communicate through a quantum channel and their goal is to compute $f(x,y)$.
Notice that in this model there is no prior entanglement between the two players.

\begin{definition}{\bf{(Quantum weak nondeterminism)}}\label{weak}
We say that such a quantum protocol is a weakly nondeterministic protocol for $f$ if,
for each $(x,y)\in X\times Y$, the following holds:\vspace{-1mm}
\begin{enumerate}
\item[(i)]
if $f(x,y)=1$ then there is a classical proof such that the protocol outputs $1$ with probability 1, \vspace{-3mm}
\item[(ii)]
if $f(x,y)=0$ then, for all classical proofs, the protocol outputs $0$ with probability 1.
\end{enumerate}
\end{definition}\vspace{-1mm}

\noindent Similarly to the classical case, the quantum weakly nondeterministic communication
complexity of $f$ is the minimum, over all the quantum weakly nondeterministic
protocols computing $f$, of the number of qubits exchanged between Alice and Bob on the 
worst-case instance and the worst proof. 
We are thus considering the worst case complexity of exact quantum protocols receiving classical proofs. 

As explained in the introduction of this paper, a stronger definition of quantum nondeterministic protocols 
can be given \cite{Massar+01,deWolfSIAMJC03}, corresponding to probabilistic protocols using quantum messages that output 1 with positive 
probability if and only if $f(x,y)=1$.
The main reasons why we think studying the power of quantum protocols resulting from Definition \ref{weak}
is meaningful is that, first, this definition corresponds to the original version of classical nondeterminism, 
based on the notion of proof, and, second, we believe quantum strongly nondeterministic
protocols are in a way too powerful to be ``fairly'' compared with classical nondeterministic protocols.
Let us give a simple example that illustrates the latter point.
The non-equality function is the function $NEQ_n:\{0,1\}^n\times\{0,1\}^n\to \{0,1\}$ such that $NEQ_n(x,y)=1$ if $x\neq y$
and  $NEQ_n(x,y)=0$ if $x=y$.
Massar, Bacon, Cerf and Cleve \cite{Massar+01} have shown a quantum strongly nondeterministic protocol for 
$NEQ_n$ using exactly one quantum bit (qubit) of communication. In comparison, it is well known that 
$N(NEQ_n)=\Theta(\log n)$ (see for example \cite{Hromkovic97}). 
We explain their simple protocol, which shows the power of quantum strong nondeterminism.
Alice sees its input $x$ as an integer in $\Set{0}{2^n-1}$, prepares the state 
$$\frac{1}{\sqrt{2}}(\cos\left(\frac{x\pi}{2^n}\right)\ket{0}+\sin\left(\frac{x\pi}{2^n}\right)\ket{1})$$ and sends it to Bob.
Bob rotates it by the angle of $-y\pi/2^n$, obtaining the state
$$\frac{1}{\sqrt{2}}(\cos\left(\frac{(x-y)\pi}{2^n}\right)\ket{0}+\sin\left(\frac{(x-y)\pi}{2^n}\right)\ket{1}).$$
Measuring this state gives $1$ with positive probability if $x\neq y$.
In the case $x=y$, then the probability of measuring $1$ is 0. The
quantum communication protocol that does the above state manipulations, measures
the final state and outputs the outcome of the measurement is thus a quantum
strongly nondeterministic communication protocol for $NEQ_n$ using only one qubit
of communication, in a way incomparable with classical nondeterminism.

\subsection{Our total function}
We now define the communication problem $HEQ_{k,k'}$ (Hadamard Equality)
that is used to show the separation of quantum weakly nondeterministic and classical nondeterministic
communication complexity.\vspace{4mm}

$\phantom{aa}${\bf Hadamard Equality} $(\mathbf{HEQ_{k,k'}},\textrm{ for } k,k'\ge 1)$\\
\indent $\phantom{aa}$Alice's input: a vector $\mathbf{a}=(a_1,\ldots,a_{2^k-1})$ in $\Set{0}{2^{k'}-1}^{2^k-1}$\\
\indent $\phantom{aa}$Bob's input: $\phantom{e}$a vector $\mathbf{b}=(b_1,\ldots,b_{2^k-1})$ in $\Set{0}{2^{k'}-1}^{2^k-1}$\\
\indent $\phantom{aa}$output: $\phantom{lalala}$$0$ if 
$(0,\delta(a_1, b_1),\ldots,\delta(a_{2^k-1}, b_{2^k-1}))\in \mathscr{H}_k\backslash\{(0,\ldots,0)\}$\\
\indent   $\phantom{aa}$$\phantom{lalalalallaal}\:\:$     $1$ else\vspace{5mm}

\noindent 
Notice that, for any $a$ and $b\in\Set{0}{2^{k'}-1}$, we have $\delta(a,b)=0$ if and only if 
$a=b$. Thus the problem 
$HEQ_{k,k'}$ can be seen as a two-leveled string equality problem: Intuitively, the hard case is for Alice and Bob to check whether
$$(0,\delta(a_1, b_1),\ldots,\delta(a_{2^k-1}, b_{2^k-1}))=(0,\ldots,0)$$ and, to do this,
they have to check whether $\delta(a_i, b_i)=0$ for sufficiently many values of $i$ (actually, at least $k$ different values). 
The point is that the nondeterministic communication complexity of testing the equality of two integers of $k'$ bits 
is $\Theta(k')$. Thus, intuitively, the classical nondeterministic communication complexity of
$HEQ_{k,k'}$ is $\Omega(kk')$. 
We will, in section \ref{section:lowerbound}, prove that when $k'$ is sufficient large,
this intuition is correct.

To our knowledge, the function $HEQ_{k,k'}$ has never been considered before, but
the case $k'=1$ is similar to a property testing problem considered by Buhrman, Fortnow,
Newman and R\"ohrig \cite{Buhrman+SODA03} in the
framework of query complexity.
The original (promise) problem in \cite{Buhrman+SODA03} is, for a fixed subset $A_k$ of $\mathscr{H}_k$, to decide
whether a string $x$ is in $A_k$ or the Hamming distance between $x$ and any string of $A_k$ is sufficiently large, by querying
as few bits of $x$ as possible.  By setting $A_k=\mathscr{H}_k\backslash\{(0,\ldots,0)\}$, and replacing
``sufficiently large'' by ``positive'', we obtain a definition similar to $HEQ_{k,1}$.
However, as far as
communication complexity is concerned, the results in \cite{Buhrman+SODA03} do not imply any separation
of classical nondeterminism and quantum weak nondeterminism. 

\section{Quantum Upper Bound}\label{section:upperbound}
In this section, we present an efficient quantum weakly nondeterministic protocol
for $HEQ_{k,k'}$. 

We first prove the following lemma, which restates, in our notations, a well-known property
of the Hadamard code.

\begin{lemma}\label{lemma:condition}
Let $\mathbf{x}=(x_0,x_1,\ldots,x_{2^k-1})$ be a vector in $\{0,1\}^{2^k}$ such
that $x_0=0$. Then the following two assertions are equivalent.
\begin{enumerate}
\item[1.]
$\mathbf{x}\in \mathscr{H}_k$;
\item[2.]
For all the indexes $i$ in $S_k$, the following holds: $x_i=x_{[i]} \oplus x_{i-[i]}$.
\end{enumerate}
\end{lemma}
\begin{proof}
Take a vector $\mathbf{x}\in \mathscr{H}_k$ and an integer $i$ in $S_k$.
From the definition of the Hadamard code, there exists a vector $\mathbf{w}\in\{0,1\}^k$
such that $x_i=\mathbf{w}\cdot\mathbf{i}$, $x_{[i]}=\mathbf{w}\cdot\mathbf{i'}$
and $x_{i-[i]}=\mathbf{w}\cdot\mathbf{i''}$, with $i'=[i]$ and $i''=i-[i]$. Then
$x_{[i]} \oplus x_{i-[i]}=\mathbf{w}\cdot(\mathbf{i'}\oplus\mathbf{i''})
=\mathbf{w}\cdot\mathbf{i}$
from the definition of $[i]$. Thus assertion 2 holds.
Now we prove that there are at most $2^{k}$ vectors in $\{0,1\}^{2^k}$ satisfying assertion 2.
Since $\abs{\mathscr{H}_k}=2^{k}$, this will prove the lemma.
Take two vectors $\mathbf{x}$ and $\mathbf{x'}$ such that
$x_0=x'_0=0$ and $x_{2^l}=x'_{2^l}$ for all $l\in\{0,\ldots,k-1\}$.
If $\mathbf{x}$ and $\mathbf{x'}$ both satisfy assertion 2 then the other bits 
are uniquely determined and thus, necessarily, $\mathbf{x}=\mathbf{x'}$. 
This implies that we can 
construct at most $2^{k}$ different vectors satisfying assertion 2. 
\end{proof}

We then present the main result of this section.
\noindent\begin{theorem}\label{theorem:comupperbound}
For any positive integers $k$ and $k'$, there exists a quantum weakly nondeterministic protocol 
using less than $3(k+k')$ qubits of communication that computes the function $HEQ_{k,k'}$.
\end{theorem}
\begin{proof}
We describe our quantum protocol, which is actually a generalization of (a modified version of) the quantum query protocol in
\cite{Buhrman+SODA03}.
Suppose that the inputs are $\mathbf{a}=(a_1,\ldots,a_{2^k-1})$, $\mathbf{b}=(b_1,\ldots,b_{2^k-1})$
and that $(\mathbf{a},\mathbf{b})$ is a 1-instance of $HEQ_{k,k'}$.
This means that one of the two following cases holds:
\begin{enumerate}
\item[(i)]
$(0,\delta(a_1,b_1),\ldots,\delta(a_{2^k-1},b_{2^k-1}))\notin \mathscr{H}_k$; or
\item[(ii)]
$(0,\delta(a_1,b_1),\ldots,\delta(a_{2^k-1},b_{2^k-1}))=(0,\ldots,0)$.
\end{enumerate}
Alice first guesses which case holds. If (i) really holds then, from Lemma \ref{lemma:condition}, there
exists an integer $j\in S_k$ such that $\delta(a_j,b_j)\neq\delta(a_{[j]},b_{[j]}) 
\oplus \delta(a_{j-[j]},b_{j-[j]})$. 
Alice guesses this index $j$,
sends the value of her guess $j$ and the three integers 
$a_j$, $a_{[j]}$ and $a_{j-[j]}$ (using a classical message). 
Bob then checks whether 
$\delta(a_j,b_j)\neq\delta(a_{[j]},b_{[j]}) 
\oplus \delta(a_{j-[j]},b_{i-[j]})$,
outputs $1$ if it holds, and $0$ else.

Now suppose that Alice guessed that (ii) holds.
Alice then creates and sends Bob the following state.
\begin{displaymath}
\frac{1}{\sqrt{2^k}}\sum_{m=0}^{2^k-1}\ket{m}\ket{a_m},
\end{displaymath}
where the first register consists in $k$ qubits and the second register $k'$ qubits. Here,
we use the convention $a_0=0$.
Bob applies the following unitary transform on the state he received:
\begin{displaymath}
\ket{m}\ket{r}\mapsto (-1)^{\delta(r,b_m)}\ket{m}\ket{r},
\end{displaymath}
for all $m\in\Set{0}{2^{k}-1}$  and $r\in\Set{0}{2^{k'}-1}$,
with the convention $b_0=0$. He then
sends back the resulting state to Alice.
Alice now performs the unitary transform
\begin{displaymath}
\ket{m}\ket{r}\mapsto \ket{m}\ket{r\oplus a_m}
\end{displaymath}
for any  $m\in\Set{0}{2^k-1}$ and $r\in\Set{0}{2^{k'}-1}$ (here $r\oplus a_m$ denote the bitwise parity of the binary encodings of
$r$ and $a_m$).
The resulting state is 
$$
\frac{1}{\sqrt{2^k}}\sum_{m=0}^{2^k-1}(-1)^{\delta(a_m, b_m)}\ket{m}\ket{0}.
$$
From now, it is simply Bernstein-Vazirani algorithm  \cite{Bernstein+SIAMJC97} (or Deutsch-Jozsa algorithm \cite{Deutsch+92}).
Alice applies an Hadamard transform on each of the $k$ qubits of the first register and measures the 
first register of the resulting state
in the computational basis, outputs 1 if the result is $0$ and outputs 0 else.
If (ii) really holds, 
the state before the measurement being $\ket{0}\ket{0}$, her
measurement result is necessarily $0$. She then
outputs $1$ without error. For any 1-instance of $HEQ_{k,k'}$,
there is thus a guess that can be verified with probability $1$ by this protocol.

Now consider the behavior of this protocol on a 0-instance, i.~e.~an instance such that
$(0,\delta(a_1,b_1),\ldots,\delta(a_{2^k-1},b_{2^k-1}))\in \mathscr{H}_k\backslash\{(0,\ldots,0)\}$.
If Alice guesses that the case (i) holds, then, from Lemma \ref{lemma:condition}, the checking procedure
always outputs $0$. If Alice guesses that the case (ii) holds, then at the end of the checking
procedure, before doing the measurement, the state will be $\ket{c}\ket{0}$ for
some $c\in\Set{1}{2^{k}-1}$. 
Measuring this state will give $c$ which is different from $0$. 
Thus the checking procedure outputs 0 with probability 1, whatever Alice's guesses are.
We conclude that the above protocol is correct on 0-instances as well.
\end{proof}

\section{Classical Lower Bound}\label{section:lowerbound}
First, notice that there exists a nondeterministic classical
protocol for $HEQ_{k,k'}$ using $O(kk')$ communication bits.
The protocol is similar to the quantum protocol of Theorem \ref{theorem:comupperbound}, but,
when Alice guesses that $(0,\delta(a_1,b_1),\ldots,\delta(a_{2^k-1},b_{2^k-1}))=(0,\ldots,0)$,
she sends the $k$ integers $a_{2^s}$, for all $s\in\Set{0}{k-1}$, instead of sending the state
$\frac{1}{\sqrt{2^k}}\sum_{m=0}^{2^k-1}\ket{m}\ket{a_m}$.
Bob then outputs 1 if and only if $\delta(a_{2^s},b_{2^s})=0$ for all these integers $s$.
The objective of this section is to show that this protocol is basically optimal.

The proof of the lower bound is based on the following strong result.
\begin{theorem}\label{mainproposition}
Let $k$ and $k'$ be two positive integers such that $k\ge 3$ and $k'\ge k$.
Consider any subset $A\subseteq \Set{0}{2^{k'}-1}^{2^k-1}$ such that, 
for any two elements $\mathbf{a}=(a_1,\ldots,a_{2^k-1})$ and $\mathbf{b}=(b_1,\ldots,b_{2^k-1})$ of $A$, the following condition holds.
\begin{equation}\label{condition}
\left\{
\begin{array}{ll}
(0,\delta(a_1,b_1),\ldots,\delta(a_{2^k-1},b_{2^k-1}))=
(0,\dots,0)&\:\: if \:\:\mathbf{a}=\mathbf{b}\\
(0,\delta(a_1,b_1),\ldots,\delta(a_{2^k-1},b_{2^k-1}))\notin\mathscr{H}_k&\:\: if \:\:\mathbf{a}\neq\mathbf{b}
\end{array}\:\:\:\right.
\end{equation}
Then $A$ necessarily satisfies
$$\abs{A}\le 2^{k'2^k-k(k'-k-1)}.$$
\end{theorem}
\begin{proof}
Our proof is inspired by a new proof by Babai, Snevily and Wilson \cite{Babai+95} 
of a result by Frankl~\cite{Frankl86}, itself generalizing a result by Delsarte~\cite{Delsarte73,Delsarte74}, that
gives an upper bound on the size of any code in function of the cardinality of the set of Hamming distances that occurs 
between two distinct codewords (but these results are fundamentally different from what we need to prove our upper bound).

Denote $\mathscr{A}=\Set{0}{2^{k'}-1}^{2^k-1}$, and consider any subset $A\subseteq\mathscr{A}$ such that any two elements
$\mathbf{a}$ and $\mathbf{b}$ satisfies the condition (\ref{condition}).
For each $a\in\Set{0}{2^{k'}-1}$, consider the polynomial $\varepsilon_a$ over the field of rational numbers defined as follows. 
$$\varepsilon_a(X)=1-
\frac{X}{a}\:\frac{X-1}{a-1}\:\cdots\:\frac{X-(a-1)}{1}\:\frac{X-(a+1)}{-1}\:
\frac{X-(a+2)}{-2}\:\cdots\:\frac{X-(2^{k'}-1)}{a-(2^{k'}-1)}\:.$$
Notice that $\varepsilon_a(b)=\delta(a,b)$ for any $a$ and $b$ in $\Set{0}{2^{k'}-1}$.
Now, given a vector $\mathbf{a}=(a_1,\ldots,a_{2^k-1})$ in $\mathscr{A}$, we define the multivariate polynomial
\begin{displaymath}
f_\mathbf{a}(\mathbf{X})=f_{\mathbf{a}}(X_1,\ldots,X_{2^k-1})=
\prod_{i\in S_k}
\big(1-\varepsilon_{a_i}(X_i)-\varepsilon_{a_{[i]}}(X_{[i]})-\varepsilon_{a_{i-[i]}}(X_{i-[i]})\big)\:.
\end{displaymath}
The polynomial $f_\mathbf{a}$ has the property
that any monomial it contains has as most 
$\abs{S_k}=2^k-k-1$ distinct indeterminates $X_j$ in it. For each $f_{\mathbf{a}}$,
we construct a new polynomial as follows:
for each variable $X_j$ appearing in $f_{\mathbf{a}}$ with an exponent $e>2^{k'}-1$, we replace
$X_j^e$ by $X_j^e$ reduced modulo $X_j(X_j-1)\ldots(X_j-(2^{k'}-1))$. 
Call $f'_{\mathbf{a}}$ the new polynomial. 
Notice that, as functions over the rationals, $f_{\mathbf{a}}$ and $f'_{\mathbf{a}}$ have the same values over $\mathscr{A}$.
As a function, each $f'_\mathbf{a}$ is in the span of all the $\sum_{i=0}^{2^k-k-1}(2^{k'}-1)^i\binom{2^k-1}{i}$ monomial functions in which at most $2^k-k-1$ 
distinct variables enter and such that the exponent of each variable is at most $2^{k'}-1$. 

From the hypothesis on $A$,
Lemma \ref{lemma:condition} implies that the following holds for all $\mathbf{a}$ and $\mathbf{b}$ in $A$.
$$f'_{\mathbf{a}}(\mathbf{b})=f_{\mathbf{a}}(\mathbf{b})\equiv
\left\{\begin{array}{ll}
1\:\:\textrm{mod } 2& \textrm{ if } \mathbf{a} = \mathbf{b} \\
0\:\:\textrm{mod } 2 & \textrm{ if } \mathbf{a}\neq\mathbf{b} 
\end{array}\right. $$
We now show that this implies that the $\abs{A}$ functions $f'_{\mathbf{a}}$ for $\mathbf{a}\in A$ are linearly independent
over the rationals. Take $\abs{A}$ rationals $\lambda_{\mathbf{a}}$ such that $\sum_{\mathbf{a}\in A}\lambda_{\mathbf{a}}f'_{\mathbf{a}}=\mathbf{0}$.
Without loss of generality, we can actually consider that the $\lambda_{\mathbf{a}}$ are integers.
The evaluation of the two sides of this expression at the point $\mathbf{b}$ gives $\lambda_{\mathbf{b}}\equiv 0\:\textrm{mod } 2$.
Thus, necessarily, $\lambda_{\mathbf{a}}\equiv 0\:\textrm{mod } 2$ for all $\mathbf{a}\in A$.
Suppose that the $\lambda_{\mathbf{a}}$ are not all zero and
denote $\Lambda_i=\{\mathbf{a}\in A\textrm{ such that }\lambda_{\mathbf{a}}\neq 0 \textrm{ and } 2^i\vert\lambda_{\mathbf{a}}\}$
for $i$ ranging from 1 to $r$, where $r$ is the greatest integer such that $2^r$ appears in the prime power decomposition of some 
$\lambda_{\mathbf{a}}$.
Evaluating, for increasing $i$ from 1 to $r$, the functions $\sum_{\mathbf{a}\in\Lambda_i}(\lambda_{\mathbf{a}}/2^i)f'_{\mathbf{a}}$  
gives that $\Lambda_{1}=\emptyset$. 
Thus $\lambda_{\mathbf{a}}=0$ for all $\mathbf{a}\in A$.

The fact that the $\abs{A}$ functions $f'_{\mathbf{a}}$ 
are linearly independent over the rationals 
implies that
\begin{eqnarray}
\abs{A}&\le& \sum_{i=0}^{2^k-k-1}(2^{k'}-1)^i\binom{2^k-1}{i}\label{eq4.1}\\
&\le& \sum_{i=0}^{2^k-k}(2^{k'})^i\binom{2^k}{i}\:\:.\label{eq4.2}
\end{eqnarray}
We now show an upper bound for this expression.\vspace{2mm}

\begin{lemma}\label{lemma:upperbound}
Let $k$ and $k'$ be positive integers such that $k\ge 3$ and $k'\ge k$. Then
\begin{displaymath}\label{equation:ineq}
\sum_{i=0}^{2^k-k}(2^{k'})^i\binom{2^k}{i}\le 2^{k'2^k-k(k'-k-1)}.
\end{displaymath}
\end{lemma}
{\bf Proof of Lemma \ref{lemma:upperbound}.}
First notice that, in the case $k'\ge k$, the function $$h:j\mapsto (2^{k'})^j\binom{2^k}{j}$$ is an increasing function over $\Set{0}{2^k}$:
For any $i\in\Set{0}{2^k-1}$, we have $h(i+1)/h(i)=2^{k'}(2^k-i)/(i+1)\ge 2^{k'}2^{-k}\ge 1$.
We can now give the following upper bound.
\begin{eqnarray*}
\sum_{i=0}^{2^k-k}(2^{k'})^i\binom{2^k}{i}&\le& 2^k\max_{i\in\Set{0}{2^k-k}}\left((2^{k'})^i\binom{2^k}{i}\right)\\
&=&  2^k(2^{k'})^{2^k-k}\binom{2^k}{2^k-k}=2^k(2^{k'})^{2^k-k}\binom{2^k}{k}.
\end{eqnarray*}
Using the standard fact $\binom{2^k}{k}\le (e2^k/k)^k$, where $e$ is the Euler constant,
we obtain, for $k\ge 3$,
\begin{eqnarray*}
\sum_{i=0}^{2^k-k}(2^{k'})^i\binom{2^k}{i}&\le& 2^k(2^{k'})^{2^k-k}\left(2^k\right)^k\\
&=& 2^{k'2^k-k(k'-k-1)}.\:\:\:\:\:
\Box
\end{eqnarray*}\vspace{2mm}
Using Lemma \ref{lemma:upperbound}, we obtain the claimed upper bound on the size of $A$. This concludes the proof of Theorem \ref{mainproposition}.
\end{proof}

We are now ready to prove the lower bound on the classical nondeterministic complexity of $HEQ_{k,k'}$.
\begin{theorem}\label{theorem:comlowerbound}
Let $k$ and $k'$ be two positive integers such that $k\ge 3$ and $k'\ge k$.
Then $$N(HEQ_{k,k'})\ge k(k'-k)-(k+k').$$
\end{theorem}
\begin{proof}
Denote again $\mathscr{A}=\Set{0}{2^{k'}-1}^{2^k-1}$.
Notice that for any $\mathbf{a}\in \mathscr{A}$, $(\mathbf{a},\mathbf{a})$ is a 1-instance of $HEQ_{k,k'}$.
We will show a lower bound on the number of 1-monochromatic (for $HEQ_{k,k'}$) rectangles of 
$\mathscr{A}\times\mathscr{A}$  
necessary to cover $\{(\mathbf{a},\mathbf{a})\:\vert\:\mathbf{a}\in\mathscr{A}\}$. Here covering means
that the union of the rectangles has only to include $\{(\mathbf{a},\mathbf{a})\:\vert\:\mathbf{a}\in\mathscr{A}\}$.
Such a lower bound obviously implies a lower bound on the number of 1-monochromatic rectangles necessary
to cover all the 1-instances of $HEQ_{k,k'}$.
Any 1-monochromatic rectangle of a covering of $\{(\mathbf{a},\mathbf{a})\:\vert\:\mathbf{a}\in\mathscr{A}\}$ can be 
considered, without loss of generality, to be of the form $A\times A$ for some subset $A\subseteq\mathscr{A}$. 
By the definition of a 1-monochromatic rectangle, for each $\mathbf{a}=(a_1,\ldots,a_{2^k-1})$
and $\mathbf{b}=(b_1,\ldots,b_{2^k-1})$ in $A$ the following must hold:
\begin{enumerate}
\item[1.]
$(0,\delta(a_1,b_1),\ldots,\delta(a_{2^k-1},b_{2^k-1}))=(0,\dots,0)\:\:$ if $\:\:\mathbf{a}=\mathbf{b}$;
\item[2.]
$(0,\delta(a_1,b_1),\ldots,\delta(a_{2^k-1},b_{2^k-1}))\notin\mathscr{H}_k\:\:$ if $\:\:\mathbf{a}\neq\mathbf{b}$.
\end{enumerate}

Then, 
even for the largest 1-monochromatic rectangle of the form $A\times A$, 
from Theorem \ref{mainproposition} we have $\abs{A}\le 2^{k'2^k-k(k'-k-1)}$. 
This implies that at least 
$$\frac{(2^{k'})^{2^k-1}}{\abs{A}}\ge 2^{kk'-k^2-k-k'}\:$$
1-monochromatic rectangles are necessary to cover $\{(\mathbf{a},\mathbf{a})\:\vert\:\mathbf{a}\in\mathscr{A}\}$.
The nondeterministic complexity of $HEQ_{k,k'}$ is thus, using Fact \ref{fact1}, at least $kk'-k^2-k-k'$. 
\end{proof}

This theorem implies the quadratic separation, as stated in the next corollary.
\begin{corollary}
There is a quadratic separation of
quantum weakly nondeterministic and classical nondeterministic communication complexity. 
\end{corollary}
\begin{proof}
By considering for example $HEQ_{k,2k}$, for which the quantum weakly nondeterministic
communication complexity is, from Theorem \ref{theorem:comupperbound}, $O(k)$ and
the classical nondeterministic communication complexity is, from Theorem \ref{theorem:comlowerbound},
$\Omega(k^2)$.
\end{proof}

\section{Discussion and Open Problems}\label{section:open}
Although we conjecture that even for arbitrary $k'$, 
the classical nondeterministic communication complexity of $HEQ(k,k')$ is $\Omega(kk')$,
it is not possible to prove this fact using
the same technique. Indeed, equation (\ref{eq4.2}) is a relatively tight
approximation of (\ref{eq4.1}) and
$$
\sum_{i=0}^{2^k-k}(2^{k'})^i\binom{2^k}{i}\ge (2^{k'})^{2^k-k}\binom{2^k}{k}
\ge 2^{k'2^k-kk'+k^2-k\log_2 k},
$$
which cannot be $2^{k'2^k-\Omega(kk')}$ when $k'$ is small with respect to $k$.

The main open problem is whether a separation larger than quadratic can be found between
classical nondeterministic and quantum weakly nondeterministic communication complexity
for a total function. Is an exponential gap achievable? 
It may indeed be the case that, for total functions, the largest gap achievable is polynomial
and, possibly, quadratic.



\end{document}